\documentclass[english,amsmath,amssymb,superscriptaddress,twocolumn,prb]{revtex4}
\usepackage{graphicx}
\usepackage{dcolumn}
\usepackage{bm}
\usepackage{subfigure}
\usepackage{float}
\usepackage{xcolor}

\begin{document}

\title{Monte Carlo Algorithm for Free Energy Calculation}

\author{Sheng Bi}
\email{ruc.bs.plu@hotmail.com}
\affiliation{Department of Physics, Renmin University of China, 100872 Beijing, China}
\affiliation{Beijing Key Laboratory of Opto-electronic Functional Materials and Micro-nano Devices (Renmin University of China)}

\author{Ning-Hua Tong}
\email{nhtong@ruc.edu.cn}
\affiliation{Department of Physics, Renmin University of China, 100872 Beijing, China}
\affiliation{Beijing Key Laboratory of Opto-electronic Functional Materials and Micro-nano Devices (Renmin University of China)}

\date{\today}

\begin{abstract}
We propose a new Monte Carlo
algorithm for the free energy calculation based on configuration space sampling. Upward or downward temperature scan can be used to produce $F(T)$. We implement this algorithm for Ising model on square lattice and on triangular lattice. Comparison with the exact free energy shows an excellent agreement. We analyse the
properties of this algorithm and compare it with Wang-Landau algorithm which samples in energy space. This method is applicable to general classical statistical models. The possibility of extending it to quantum systems is discussed.
\end{abstract}

\maketitle
\section{Introduction}

Monte Carlo (MC) simulation is one of the most important numerical 
methods for solving statistical problems in fields such as
chemistry, biology and physics. In condensed matter physics,
MC is extensively used to study the properties of many
statistical models, phase transitions, and quantum many-body
 systems [\onlinecite{Landau1}]. Often, besides the expectation
values of certain physical quantities, one also needs the
 free energy of the system in thermal equilibrium. Calculation of free energy is a difficult problem for traditional Metropolis algorithm since partition function plays the role of the normalization constant in the thermal probability density distribution of an ensemble.

In the past three decades, various MC algorithms have been
proposed to do the free energy calculation for statistical Hamiltonians. For classical systems, the frequently used methods are the histogram reweighting method [\onlinecite{Ferrenberg1}], multiple histogram [\onlinecite{Ferrenberg2, Valleau1,Alves1}], transition matrix MC [\onlinecite{Wang1}],  entropic sampling[\onlinecite{Lee}], flat histogram method [\onlinecite{Berg}], and the Wang-Landau method [\onlinecite{Wang2}]. Recently Wang-Landau method has been improved in different aspects [\onlinecite{Belardinelli1},\onlinecite{Vogel}]. All these methods have their respective advantages and disadvantages. For examples, the histogram reweighting method produces the energy
histogram $P_{T_0}(E)$ at a given temperature $T_0$ and employs the
reweighting method to recover the distribution at a different
temperature. As is shown below, usually the canonical distribution
$P_{T}(E)$ is sharply peaked around $\langle E\rangle_{T}$ which is $T$-dependent.
Thus the error of free energy becomes large when $|T-T_0|$ is large. For both the entropic sampling and the Wang-Landau method, the ensemble is created in the
energy space instead of in the configuration space. This helps to
obtain the density of states efficiently, but it is less convenient if 
one hopes to study other physical quantities in a single simulation,
especially when these quantities are not simple functions of energy $E$, such as the correlation function. 

For quantum systems, the calculation of free energy is also a pertinent but harder problem. In this regard, the Wang-Landau algorithm has been combined with statistical series expansion method to calculate the free energy of quantum Hamiltonians such as the Heisenberg model [\onlinecite{Troyer}]. The idea of flat histogram has been applied to the diagrammatic MC method to improve the long imaginary-time results [\onlinecite{Diamantis}],
and to calculate the grand potential of a cluster system with electron bath [\onlinecite{Li,Gull}]. However, considering that the flat-histogram method or Wang-Landau algorithm have not been implemented under the path integral(PI) quantum Monte Carlo (QMC) methods such as the determinantal QMC [\onlinecite{Suzuki,Hirsch1}] and the continuous-time PI QMC method [\onlinecite{Beard}], which are based on the Metropolis sampling in configuration space, it is still desirable to develop a free energy calculation method which can calculate free energy by the configuration space sampling.

 It is the purpose of this paper to propose such a new MC algorithm that can calculate the free energy using the configuration-based sampling algorithm. The price that we have to pay is a sequential scan from low to high temperatures, or vice versa. We demonstrate the implementation of this algorithm using the Ising model on square as well as on triangular lattice, for which the exact solutions are known. Comparison with Wang-Landau method shows that both efficiency and accuracy of this method are satisfactory. The additional advantage of this method is that it is based on Metropolis algorithm and hence in principle it can be extended to quantum systems within determinantal or path integral methods.

\section{Method and Results}

In this part, we demonstrate the implementation of our method and analyse its features using the two dimensional Ising model on a square lattice. For comparison purposes, here we use the equivalent Hamiltonian of the two-state Potts model
\begin{equation}
  H=-J \sum_{\langle i,j \rangle} \delta_{s_i,s_j} .
\end{equation}
Here, $s_i$ is the spin degrees of freedom on site $i$ and it takes
integer values from $0$ to $1$. $\delta_{s_i,s_j}=0$ if $s_i \neq
s_j$ and $\delta_{s_i,s_j}=1$ if $s_i = s_j$. The summation is for pairs of nearest neighbour sites on a square lattice with $N \times N$ geometry. This model has been studied extensively as a basic statistical model [\onlinecite{Wu}]. Its exact critical transition temperature is $k_{B}T_c = 2J/\text{ln}(1+\sqrt{2})$. In the following we use $J=1$ as the energy unit and set the Boltzmann constant $k_{B}=1$ for convenience.

One of the widely used MC algorithms for studying classical
statistical models is the Metropolis sampling in configuration space [\onlinecite{Metropolis}]. In this algorithm, one begins by choosing a random spin configuration $S_0$ (here we use capital letter $S$ to denote the spin configuration of the whole
lattice), and update the configuration $S_{n} \rightarrow S_{n+1}$
according to a given proposal probability $P(S_{n} \rightarrow S_{n+1})$
and an accepting probability $A(S_{n} \rightarrow S_{n+1})$. The
transition probability $T(S_{n} \rightarrow S_{n+1}) = P(S_{n}
\rightarrow S_{n+1}) A(S_{n} \rightarrow S_{n+1})$ must satisfy the
detailed balance condition $f\left(S_{n} \right)T(S_{n} \rightarrow
S_{n+1}) = f\left(S_{n+1} \right)T(S_{n+1} \rightarrow S_{n})$ to
guarantee that the resulting Markovian chain has the target
distribution $f\left( S_{n}\right)$ in the equilibrium limit.
For the thermodynamical calculation, we use the Bolzmann distribution as our
target distribution,
\begin{equation}
f \left( S_n \right) = \frac{1}{Z} e^{-\beta H( S_n )},
\end{equation}
$Z$ is the partition function $Z = \sum_{S} e^{- \beta H(S)}$. Here $\beta=1/T$ is the inverse temperature. After the Markovian chain reaches equilibrium, sampling on this chain can produce the required statistical averages.
\begin{figure}[t!]
\vspace{-1.0in}
\begin{center}
\includegraphics[width=6.8in, height=4.5in, angle=0]{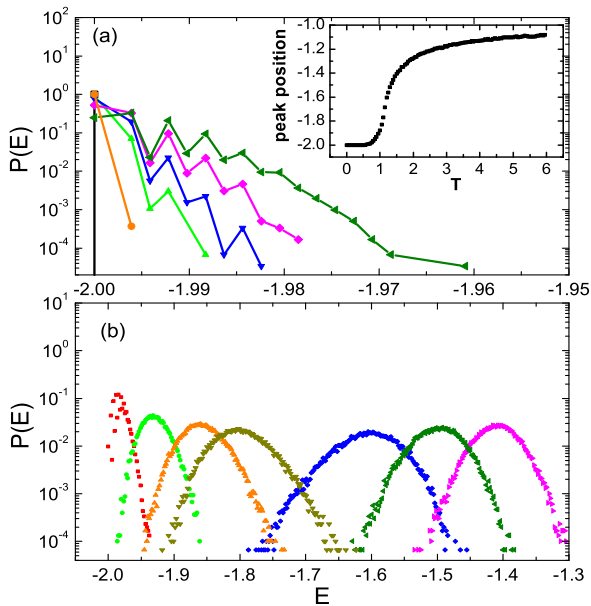}
\vspace*{-2.5cm}
\end{center}
\caption{(Color online) The normalized energy distribution probability for the Ising model on square lattice with $N=32$. (a)
From left to right, $T=0.18, 0.24, 0.42, 0.48, 0.54, 0.6$; (b) from left to right, $T=0.72, 0.9, 1.02, 1.08, 1.2, 1.32, 1.5$. Inset of (a): the peak position as function of $T$.}
\end{figure}
However, the free energy $F=-\beta \text{ln}Z$ cannot be calculated directly, because the partition function $Z$ is a normalization factor of the probability distribution $f\left( S \right)$
and hence cannot be treated as a statistical average. To calculate $F$, usually one either employs the concept of energy histogram [\onlinecite{Ferrenberg1}] within MC method, the maximum entropy method [\onlinecite{Huscroft}], or by numerical integration over the derivative of free energies [\onlinecite{Tong}]. 
 In the following, we propose a new method which combines the idea of energy histogram and the configuration-space sampling to calculate the free energy over full temperature range. 

In the Metropolis algorithm, the energy probability distribution
produced by the Markovian chain is
\begin{equation}
   p(E)=\frac{g(E)}{Z} e^{-\beta E},
\end{equation}
where $g(E)$ is the degeneracy of energy level $E$ of the given
Hamiltonian. $p(E)$ can be estimated approximately from the energy
histogram of the Markovian chain, that is,
\begin{equation}
    p(E) \approx \frac{N(E)}{m}.
\end{equation}
Here $N(E)$ is the number of spin configurations with energy $E$
and $m$ is the total number of sampled configurations
in the Markovian chain. The precision of the above estimation increases as
$m$ increases. In the limit $m \rightarrow \infty$, we get
$p(E) = g(E)e^{-\beta E}/Z = N(E)/m$,
and hence
\begin{equation}
   F(T)= -\beta \text{ln}Z = -\beta \text{ln} \left[ g(E)e^{-\beta E}\frac{m}{N(E)}  \right].
\end{equation}
In principle, $F(T)$ can be calculated from the values of $g(E)$ and $N(E)$ at any given energy $E$. For a large variety of classical Hamiltonians, the
ground state degeneracy $g(E_{g})$ is easy to obtain. For the
Ising model on square lattice, for an example, $g(E_{g})=2$. $F(T)$ is  thus in principle obtainable from the ground state energy histogram $N(E_g)$.

In practice, however, the above simple scheme does not work at arbitrary $T$ because the energy distribution $p(E)$ is sharply peaked at an energy $E(T)$ (or at many different $E$'s for some models) which is an increasing function of $T$.
In Fig.1, we show $p(E) \approx N(E)/m$ for the Ising model on 
the square lattice with $N=32$, at different temperatures. 
Here $E$ is the energy per site and $E_g=-2.0$. $p(E)$ is plotted in
logarithmic scale and it decays very fast away from the peak
position. At high temperatures, $p(E)$ is peaked at high energy and 
$p(E_g)$ is so small that it is impossible to calculate it accurately 
from $N(E_g)/m$, because at high temperatures the latter is practically zero for finite $m$. 

One way to overcome the above difficulty is to use Eq.(3) at a different temperature $T^{\prime}$,
\begin{equation}
   p^{\prime}(E)=\frac{g(E)}{Z^{\prime}} e^{-\beta^{\prime} E}.
\end{equation}
Using the fact that $g(E)$ is $T$-independent, the ratio of Eq.(6) and Eq.(3) gives
\begin{equation}
   p^{\prime}(E)= p(E) e^{-(\beta^{\prime} - \beta) E} \frac{Z}{Z^{\prime}},
\end{equation}
which can be used to evaluate the ratio of partition function at $T$ and $T^{\prime}$ as
\begin{equation}
    \frac{Z}{Z^{\prime}} =\frac{\int p^{\prime}(E) dE}{\int p(E) e^{-(\beta^{\prime} - \beta) E} dE }.
\end{equation}
This strategy works as long as for some energy $E$, $p(E)$ can be measured accurately both at $T$ and $T^{\prime}$. It is the basis for  the histogram reweighting method [\onlinecite{Ferrenberg1}] and the multiple histogram method [\onlinecite{Ferrenberg2,Valleau1,Alves1}] to produce the free energy $F$ up to a constant factor. This constant factor can be further fixed by using the infinite temperature partition function $Z(T=\infty)$ [\onlinecite{Valleau1}] or the integral of density of states $\int g(E) dE$ (Ref.~\onlinecite{Alves1}), both of which are equal to the total number of configurations.

Here, we will take a different strategy, which is based on Eq.(3) at the same temperature but at a different energy $E^{\prime}$, $p(E^{\prime}) =   g(E^{\prime})e^{-\beta E^{\prime}}/Z$. From the ratio of it and Eq.(3) one obtains
\begin{eqnarray}
    g(E^{\prime}) & =& g(E) \frac{p(E^{\prime})}{p(E)} e^{-\beta(E-E^{\prime})}    \nonumber \\
      &\approx & g(E) \frac{N(E^{\prime})}{N(E)} e^{-\beta(E-E^{\prime})} .
\end{eqnarray}
This means that the knowledge of $g(E)$ can be transferred from one energy to all $E \in W_{T}$, where $W_{T}$ is the energy window in which $N(E)/m$ is reasonably large. As shown in Fig.1, the center of this window increases with temperature $T$. 

For those Hamiltonians where the ground state degeneracy $g(E_g)$ is known, to obtain $g(E)$ at higher energies, we need to increase $T$ from a small value in such small steps that the energy windows $W_i$ and $W_{i+1}$ of adjacent temperatures $T_i$ and $T_{i+1}$ have significant overlap. Suppose one knows $g(E_i)$ for a certain energy $E_i \in W_i$. one does the MC calculation at $T_i$ to produce the histogram $N(E)/m$ for each $E \in W_{i}$. Using the known $g(E_i)$ value, $F(T_i)$ is obtained from Eq.(5) and $g(E)$ for each $E \in W_{i}$ is obtained from Eq.(9). For the next temperature $T_{i+1}$, one calculate the histogram $N(E)/m$ for each $E \in W_{i+1}$. We choose an energy $E_{i+1} \in \left( W_i \cap W_{i+1} \right)$. Using $g(E_{i+1})$ value obtained in the $T_i$ calculation and $N(E_{i+1})/m$, $F(T_{i+1})$ is obtained from Eq.(5) and $g(E)$ for each $E \in W_{i+1}$ is obtained from Eq.(9). This process goes on until the desired high temperature is reached. In this way, $g(E)$ for larger and larger energies and $F(T)$ at successively higher temperatures can be obtained. We call this scheme the upward temperature scan.

One could also start from the high temperature limit $T=\infty$, and employ the direct sampling without Boltzmann factor to produce $g(E) = Z(T=\infty) N(E)/m$ for energies $E \in W_{T=\infty}$. By decreasing $T$ from $T=\infty$ step by step, one could reach low $T$ and calculate $F(T)$ for all temperatures. This scheme is called downward temperature scan.

In the implementation of the above algorithm, an important technical issue is how to select the common energy point $E_{i+1} \in \left( W_{i} \cap W_{i+1} \right)$. In our calculation, we use the crossing energy $E_c$, which is determined by $p(E_c, T_i) = p(E_c, T_{i+1})$. It is the energy where the $p(E)$ curves of adjacent temperatures $T_i$ and $T_{i+1}$ crosses. The $E_c$ value chosen in this way has the largest value $N(E)/m$ for both $T_i$ and $T_{i+1}$, hence guarantees the optimal precision.

After MC calculation at $T=0$ ($T=\infty$) is carried out, the rest calculations are done at equal distance temperatures $T_i = T_0 + i \delta T$ ($T_i = T_0 - i \delta T$) ($i=0,1,2,...$) for upward (downward) $T$ scan. For the upward scan, we use $T_0=0.05$ and for downward scan, $T_0=6.0$. We carefully control the interval $\delta T$ to reach the optimal precision. A too large $\delta T$ will lead to a small $p(E_c)$ and relatively large error in $N(E_c)/m$. If $\delta T$ is too small, the number of temperatures required in scan from $T_0$ to the final $T$ will increase. It leads to longer calculation time and larger accumulated error in the $g(E)$ transfer. Therefore, a suitable $\delta T$ should be found by testing calculations. This issue is discussed in below (see Fig.6).

\subsection{upward temperature scan}
In our benchmark calculation for the Ising model on a square lattice, we use the cluster update scheme of Wolff [\onlinecite{Wolff}]. It has a relatively high updating speed and weak critical slowing down near the critical temperature. The free energy calculation algorithm described above can also be used with local update algorithm without modification. In Fig.1, we show the normalized energy probability distribution obtained from the Markovian chain, for $T< 0.6$ (in Fig.1(a)) and for relatively high temperatures 
$T \geq 0.6$ (in Fig.1(b)). Here $E$ is the energy per site. At low $T$, $p(T)$ has a peak at $E_{g}=-2.0$ and its width broadens with increasing $T$. While for $T \geq 0.6$, the peak position begins to
increase with temperature and its width gets saturated.
The peak position as a function of $T$ is shown in the inset of Fig.1(a). It moves from $E=-2.0$ at $T=0$ to $E=-1$ in large $T$ limit.
Note that the sharp increase occurs near $T =1.2$, close to the $T_c$ of order-disorder phase transition of this system in the thermodynamic limit [\onlinecite{Ferdinand}]. Significant overlap in the peak energy windows for adjacent temperatures $T_i$ and $T_{i+1}$ is crucial for our algorithm to work. In this work, we use a uniform temperature mesh and choose $\delta T = 0.05$ such that the overlap of the peaks are large enough to guarantee the precision. 

\begin{figure}[t!]
\vspace{0.0in}
\begin{center}
\includegraphics[width=4.8in, height=3.7in, angle=0]{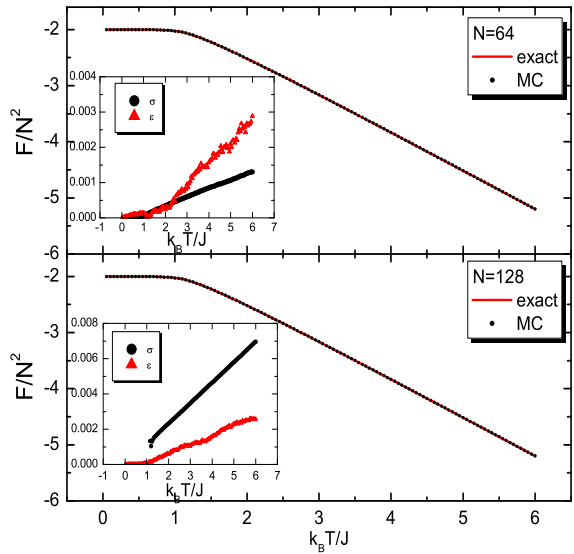}
\vspace*{-2.0cm}
\end{center}
\caption{(Color online) The free energy per site of Ising model as function
of $T$, calculated from present method with upward $T$ scan (black dots) and from the exact solution (red line) for (a) $N=64$ and (b) $N=128$, respectively. Both are obtained using $m=3 \times 10^{4}$. 
The dots and the line are indistinguishable
in the present scale. Inset: standard variance $\sigma$ (dots) and
the numerical error $\epsilon$ (triangles) of $F/N^2$. $\sigma$ is
estimated from $400$ independent Markov chains.}
\end{figure}
In Fig.2(a), the free energy per site obtained from the upward $T$ scan for $N=64$ is shown. It is compared with the exact
result [\onlinecite{Ferdinand}]. Except stated otherwise, we use $m=3 \times 10^{4}$ samples to calculate $F$ for each temperature $T_i$. With $\delta T = 0.05$, producing $F(T)$ in the range $0 < T \leq 6.0$, requires $120$ temperature values $T_i$. In Fig.2, our results are indistinguishable from the exact curve on the scale of the figure. In the inset, the standard variance $\sigma = \sqrt{\langle F^{2} \rangle - \langle F \rangle^{2}}/N^{2}$ is shown as functions of $T$.
For each temperature, $\sigma$ is measured from $400$ independent data of $F$, each of which is produced by $m=3 \times 10^4$ sampling. The actual error $\epsilon= |\langle F \rangle - F_{exc} |/N^{2}$ has same order of magnitude as $\sigma$ and it is always less than $3 \sigma$. Both quantities are small in $T<1.15J$ and increase linearly with temperature in $T> 1.15$. $T=1.15$ is close to the order-disorder transition occurring at $T_c \approx 1.181$ in the thermodynamic limit. The linear increase of error with $T$ in the disordered phase is related to the constant relative standard variance $\sigma_{Z}/Z$ of the partition function, as discussed in Fig.3 below. For the highest
temperature that we study $T=6.0$, the actual error is smaller
than $4 \times 10^{-3}$. The relative error $\epsilon/|F_{exc}|$ first increases with $T$ and then  saturates to $6 \times 10^{-4}$ in $T \geq 0.6$. 

The same comparison is made for larger lattice size $N=128$ in Fig.2(b). The results are similar to $N=64$ and the agreement between 
MC and the exact result is excellent. 
The main difference from the $N=64$ case is that the
standard variance $\sigma$ is larger than the actual error
$\epsilon$. As temperature increases, an abrupt
increase of $\sigma$ and $\epsilon$ occurs at $T \sim 1.15$ and the
linear $T$ behavior occurs at $T \gg 1.15$. These features are same as $N=64$ case.
For the relative error, a similar saturation in $\epsilon/|F_{exc}|$ 
is observed in high temperature to about $6\times 10^{-4}$.

\subsection{downward temperature scan}
In Fig.3(a), we show $F$ per site as function of $T$ obtained from the downward $T$ scan method, for the same Ising model on the square lattice. It is noted that the numerical error obtained from downward scan is of the same order as the upward scan in Fig.2. Compared to Fig.2, the $\sigma(T)$ and $\epsilon(T)$ have a sharp peak around $T_{c}$, instead of a kink. In order to understand the linear $T$-dependence of the errors in Fig.2 and its difference with Fig.3(a), in Fig.3(b), we show the relative standard variance of the partition function $\sigma_{Z}/Z$, which is related to the variance of $F$ by $\sigma_{F} = T \sigma_{Z}/Z$. It is seen that for the upward $T$ scan, $\sigma_z/Z$ surges at $T_c$ by one magnitude, from smaller than $10^{-4}$ below $T_c$ to about $10^{-3}$ above $T_c$. Further increasing temperature it does not change much. For the downward $T$ scan, $\sigma_z/Z$ is small for $T> T_c$ and surges at $T_c$ to about $10^{-3}$. It stays at this value to $T=0$. Clearly, in both cases, the transfer error of $g(E)$ from one $E$ to another is largest around $T_c$. 
As shown in the inset of Fig.1(a), the peak position in $p(E)$ curve moves fastest near $T_{c}$. For a uniform temperature mesh and constant peak width, the overlap of $p(E)$ is therefore smallest for $T_i$ and $T_{i+1}$ close to $T_c$, leading to the largest accumulation of error in $g(E)$. 
Such behavior of $\sigma_z/Z$ as a function of $T$ translates into the the linear behavior of $\sigma_F(T)$ for $T$ away from $T_c$ and the kink or the sharp peak for $T$ close to $T_c$, as shown in the insets of Fig.2 and Fig.3(a).

\begin{figure}[t!]
\vspace{-0.0in}
\begin{center}
\includegraphics[width=6.0in, height=4.7in, angle=0]{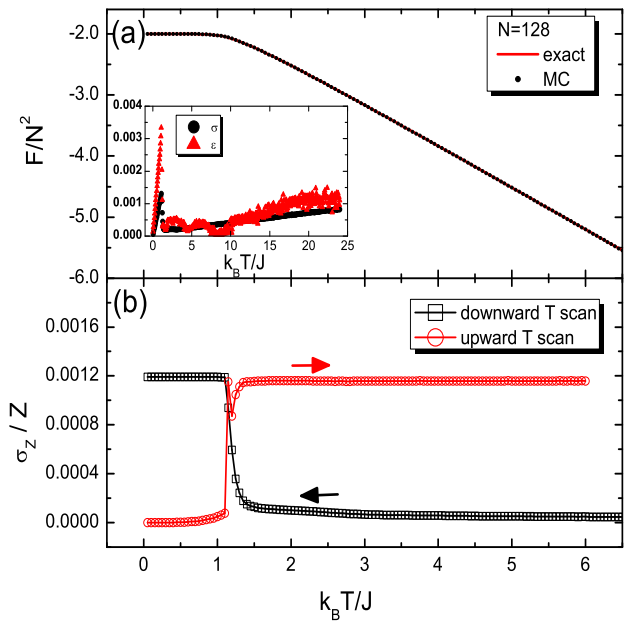}
\vspace*{-3.5cm}
\end{center}
\caption{(Color online) (a) The free energy per site of Ising model as function
of $T$, calculated from the downward $T$ scan (black dots) and from the exact solution (red line) for $N=128$ square lattice. It is obtained using $m=3 \times 10^{4}$. 
The dots and the line are indistinguishable
in the present scale. Inset: the standard variance $\sigma$ and the actual error $\epsilon$ as functions of $T$. (b) The relative standard variance $\sigma_Z/Z$ of the partition function $Z$, obtained from the upward $T$ scan (circles, right arrow) and the downward $T$ scan (squares, left arrow), respectively. They are estimated from $400$ independent Markov chains.}
\end{figure}

\subsection{Ising model on triangular lattice}
To show that our method works also for other systems, especially the frustrated system, we apply the downward temperature scan to the Ising model on a triangular lattice. This system has been solved exactly [\onlinecite{Wannier1,Houtappel1}]. The frustration induces huge low energy degeneracy and makes this system particular interesting.

In our calculation, we consider a $N \times N$ square lattice with the nearest neighbour coupling as well as the next nearest neighbour coupling in the $-xy$ direction. Both couplings are antiferromagnetic. The Hamiltonian is the same as Eq.(1), but with $J=-1.0$. Open boundary condition is used. In Fig.4(a), $F$ per site is shown as a function of $T$. Compared to the square lattice case, except for the similarity in the shape of curve, there are two important differences. One is that in the low temperature limit, $F(T)/N^2$ approaches $1.0$ instead of $-2.0$ as in the square lattice case. This shows that frustration increase the free energy of the spins. The second difference is that $F(T)$ is more smooth and there is no transition point associated with a finite temperature phase transition. This is consistent with the fact that there is no finite temperature transition in the triangular lattice Ising model [\onlinecite{Wannier1,Houtappel1}]. In the inset of Fig.4(a), the variance of $F$ decreases smoothly as $T$ is lowered and increases sharply at very low temperatures. It remains to be elucidated whether this behavior is related to the singular ground state correlation of the Ising model on a triangular lattice [\onlinecite{Wannier1}].

Due to the difference in Hamiltonian definition, it is difficult to compare our MC free energy $F(T)$ with the exact result in Ref.~\onlinecite{Wannier1}. In Fig.4(b), we show the entropy per site $S(T)/N^2$ as functions of $T$ for various lattice size $N=8, 16$ and $32$. 
It is calculated from the energy $U$ and $F$ via $S=(U-F)/T$. In the high temperature limit, $S/N^2$ approaches $\text{ln}2$ (not shown). In the zero temperature limit, our results for system $N=4, 5$ and $6$ from exact enumeration shows that $S(0)/N^2 = \text{ln}2/N^2$, being consistent with the observed two fold degeneracy in the ground state. However, for a low but finite temperature, entropy increases with $N$, as shown in Fig.4(b). For $N=32$, entropy approaches $0.315$ at $T=0.05$, the lowest temperature that we study. This value is already very close to the exact $S_{exc}(0) = 0.3383$ in the thermodynamic limit [\onlinecite{Wannier1}]. This supports the reliability of our calculation of free energy for this system. Note that for a given $N$, the error bar increases with lowering temperature decreases mainly due to the sharp increase of error in $F/N^2$ in downward T-scan. For a fixed temperature, the error bar decreases with increasing $N$ because of the $1/N^2$ factor, showing that our method is robust for large systems.

\begin{figure}[t!]
\vspace{-0.0in}
\begin{center}
\includegraphics[width=5.8in, height=4.6in, angle=0]{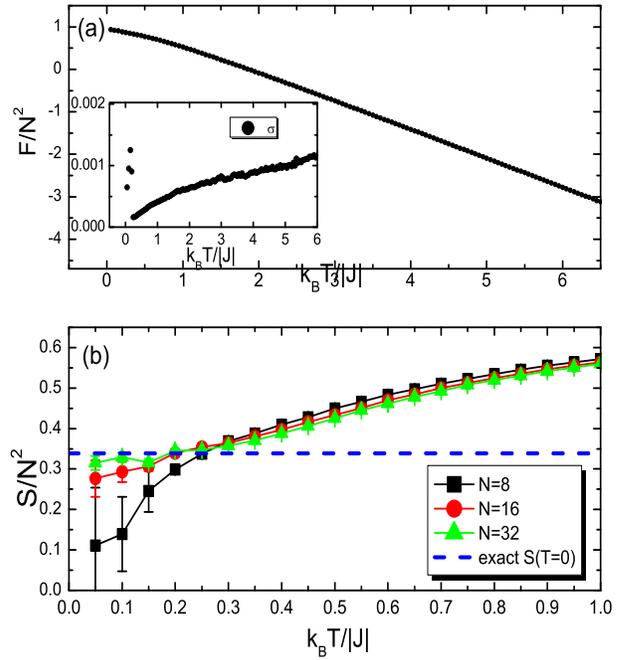}
\vspace*{-3.0cm}
\end{center}
\caption{(Color online) (a) The free energy per site of Ising model on the $32 \times 32$ triangular lattice as function of $T$, calculated from downward $T$ scan. It is obtained using $m=3 \times 10^{4}$. 
Inset: the standard variance $\sigma$ as a function of $T$. (b) The entropy per site as function of $T$ from downward $T$ scan. The dashed line is the exact value $S(T=0)=0.3383$ [\onlinecite{Wannier1}]. }
\end{figure}

\section{Discussions}

\begin{figure}[t!]
\vspace{-1.5in}
\begin{center}
\includegraphics[width=6.5in, height=4.5in, angle=0]{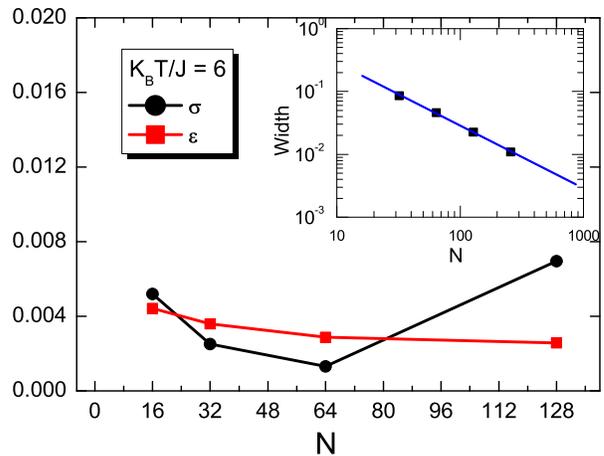}
\vspace*{-1.5cm}
\end{center}
\caption{(Color online) The standard variance $\sigma$ and the numerical error
$\epsilon$ of $F$ as functions of linear size of the lattice $N$. 
Inset: full width at half maximum of $p(E)$ curve as a function of $N$.
Other parameters are $k_{b}T/J=6.0$ and $m=3 \times 10^{4}$. $\sigma$ is estimated from $400$ independent Markov chains.}
\end{figure}
\begin{figure}[t!]
\vspace{-0.0in}
\begin{center}
\includegraphics[width=4.2in, height=3.0in, angle=0]{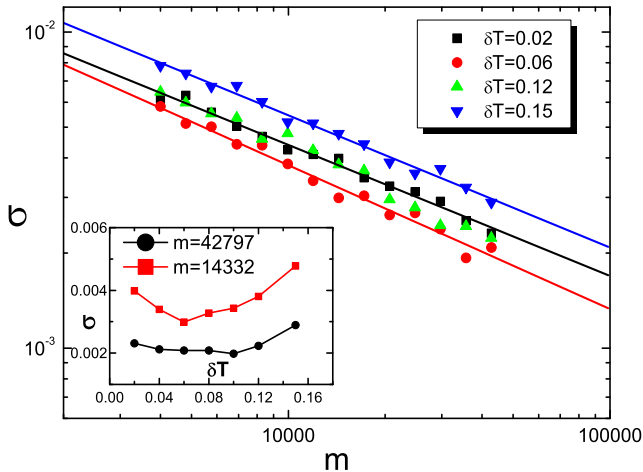}
\vspace*{-1.0cm}
\end{center}
\caption{(Color online) The standard variance $\sigma$ of $F$ at $k_{b}T/J=6.0$ as
functions of $m$ for various $\delta T$ values. $\sigma$ is estimated
from $400$ independent Markov chains for $N=32$ lattice. The solid lines
lines are power law fit of $\sigma \propto m^{x}$ with an average $x= -0.44$.
The inset shows $\sigma$ as functions of $\delta T$ for different $m$ values.}
\end{figure}
In this section, we discuss the calculation error of the free energy and compare it with the Wang-Landau's algorithm. To see the dependence of error on system size $N$, sampling number $m$, and the temperature mesh $\delta T$, we take the data from upward $T$ scan for square lattice Ising model with $J=1.0$. First, we present the $N$-dependences of $\sigma$ and $\epsilon$ in Fig.5. It is seen that for all the calculated size $ 16 \le N \le 128$, $\sigma$ and
 $\epsilon$ are on the same order, all smaller than $8 \times 10^{-3}$.
 This shows that the efficiency of our algorithm does not deteriorate 
with increasing $N$, at least for $N \le 128$. The relative 
magnitude of $\sigma$ and $\epsilon$ may vary for different $N$'s, but the actual error $\epsilon$ is always within $3 \sigma$. In the inset of Fig.5,
we show the full width at half maximum of $p(E)$ as a function of $N$ at $T=6.0$.
It is observed that the peak width scales as $1/N$. This reminds us that for very large $N$,
the $p(E)$ curves will be very sharp and we have to use a denser $T$-mesh. At $N=128$ and $\delta T=0.05$, however, the actual effective peak width is of the same order as the peak distance even in the worst case of vicinity of $T_c$ . Therefore, we can still reach similar precision as for smaller systems.
For even larger size, $\delta T$ needs to be scaled as $1/N$ to maintain the balance of peak width and peak distance, and thus maintain the calculation precision.

In Fig.6, we show the standard variance $\sigma$ as functions of sampling number $m$ for various $\delta T$ values. Our results on $N=32$ square lattice shows that $\sigma \approx m^{-x}$ with the average $x=0.44$, close to the expected $1/2$ from the central limit theorem. For fixed $m$, $\sigma$ has a weak dependence on $\delta T$ in the range $0.02 < \delta T < 0.15$. As shown in the inset of Fig.6, $\sigma$ is approximately a parabolic function of $\delta T$, with the minimum reached at a $m$-dependent $\delta T$. For all $m$ values we used, we find that the smallest error is reached at $\delta T$ values around $\delta=0.06 \sim 0.1$.

We make quantitative comparison with Wang and Landau's results [\onlinecite{Wang2}], although this method may not be the most accurate one of free energy calculation [\onlinecite{Wang1}]. We find that when scaled to the same $m$ values, our result of $F(T)$ has an error about one magnitude larger than that in Ref.\onlinecite{Wang2}. Simple optimization of our method can reduced the error significantly, such as using denser T-mesh near $T_c$ and sparser one away $T_c$, or using upward T-scan for $T < T_c$ and downward T-scan for $T> T_c$.

Unlike the Wang-Landau algorithm which generates random walks in energy space, our algorithm  samples directly in the configuration space. As a result, this method could be fitted into certain MC simulation of quantum systems. In the PI QMC[\onlinecite{Beard}] and the determinantal QMC [\onlinecite{Hirsch1}] methods, the partition function of a given quantum Hamiltonian is expressed by the summation over configurations of classical auxiliary fields. The proposed algorithm can then be used to calculate $F(T)$, using the same Markovian chain as used for evaluating general expectation values. 

One example of the application of $F(T)$ appears in the study of the Mott metal-insulator transition [\onlinecite{Gebhard}] in the half-filled Hubbard model using the dynamical mean-field theory (DMFT) [\onlinecite{Vollhardt, Kotliar}], which is exact in infinite spatial dimensions. The transition from the Fermi liquid state in small $U$ regime into the Mott insulator in large $U$ regime was found to be a special second order phase transition at $T=0$ and a first order one at $T>0$. To determine the actual transition line at $T>0$, one needs to compare the free energy of the two coexisting phases within the two spinodal lines. Within DMFT, this task is reduced to the evaluation of free energy of the effective Anderson impurity model [\onlinecite{Kotliar}]. This proves to be a difficult problem for QMC methods such as the Hirsch-Fye algorithm [\onlinecite{Hirsch2,Bluemer}]. Recently, the grand potential of the cluster problem is calculated by Wang-Landau method combined with the continuous time QMC [\onlinecite{Li}], and it is used to calculate the grand potential of lattice model within cluster dynamical mean-field theory. It is an interesting topic to apply our algorithm in various QMC methods to handle similar problems. Work in this direction is under progress.

\section{Acknowledgements}
We are grateful to the helpful discussions with Prof. You-Jin Deng.
This work is supported by 973 Program of China (2012CB921704), NSFC grant (11374362), Fundamental Research Funds for the Central 
Universities, and the Research Funds of Renmin University of China.

\end{document}